\documentclass[%
 reprint,
 amsmath,amssymb,
 aps,
]{revtex4-1}

\usepackage{graphicx}
\usepackage{dcolumn}
\usepackage{bm}

\begin{document}

\preprint{APS/123-QED}

\title{Gauge Invariance and the Quantum Metric Tensor}

\author{J. Alvarez-Jim\'enez}
\author{J. D. Vergara}%
 \email{vergara@nucleares.unam.mx}
\affiliation{Instituto de Ciencias Nucleares, Universidad Nacional Aut\'onoma de M\'exico, A. Postal 70-543 C. P. 04510, Ciudad de M\'exico, M\'exico.
}%




\date{\today}

\begin{abstract}
The  quantum metric tensor was introduced for defining the distance in the parameter space of a system. However, it is also useful for other purposes, like predicting quantum phase transitions. Due to the physical information this tensor  provides, its gauge independence sounds reasonable. More over, its original construction was made by looking for this gauge independence. The aim of this paper, however,  is to prove that the quantum metric tensor does depend on the gauge. In addition, a  real gauge invariant quantum metric tensor is introduced.  In this paper, the gauge dependence is  explicitly shown  by computing  the quantum metric tensor of the Landau problem in different gauges.  Then, a real gauge independent metric tensor is proposed and computed for the same Landau problem. Since the gauge dependence has not been observed before, the results of this paper might lead to a new study of topics that were believed to be completely understood.


\begin{description}
\item[Keywords]
Landau problem,  Quantum Metric Tensor, Gauge Dependence.
\end{description}
\end{abstract}

\maketitle


\section{\label{sec:intro}Introduction}

 The main purpose for constructing the quantum metric tensor (QMT) was to  define a  distance in the system's parameter space \cite{estreim}. That is why it is not surprising that the QMT is related to the quantum fidelity, which is also used for measuring the distance between states \cite{fideqft}. However, some studies have shown that the QMT can also be used to predict quantum phase transitions \cite{trfastmq}.
Moreover, the geodesics induced by the QMT have been useful for analysing the phases of a system that presents second order quantum phase transitions \cite{geodept}. In general, the Riemannian structure introduced by the QMT has been studied in some particular systems, see for example \cite{qmtgrstman}.

  The QMT was constructed by looking for a gauge independence \cite{estreim} and,  in fact, it was partially done. However, when we consider some kinds of  gauge transformations,  the QMT is not invariant.  Nevertheless, in current works the gauge dependence is overtly assumed.  Since this gauge dependence  has not been observed before,  we  explain its origin and propose a real gauge invariant quantum metric tensor. 

In this paper, we use the Landau problem to show the gauge dependence of  the QMT. For this reason,  in Section \ref{sec:landprob} we describe the Landau problem  in the symmetric gauge.  Section \ref{sec:qmtoflanp} shows the  QMT for one of the ground states in different gauges. While Section \ref{sec:reinqmt}  introduces a gauge independent definition of the QMT,  Section \ref{sec:gaugeinvariant}  shows the calculation of this new definition for the Landau problem. Finally,   our conclusions are written in Section \ref{sec:conclu}.

\section{\label{sec:landprob}The Landau Problem}
The Landau problem \cite{Landau} consists on a charged particle interacting with a constant and homogeneous magnetic field, $\vec{B}$.
If we consider a particle of unitary mass and  charge  $e < 0$ the Hamiltonian of the system is given by 
\begin{equation} \label{eq:tham}
H = \frac{1}{2 } \left( \vec{p} - \dfrac{e}{c} \vec{A} \right)^2, 
\end{equation}
where $\vec{A}$ is the vector potential, such that $\vec{B} = \nabla \times \vec{A}$. If we assume that the magnetic field points in $z$ direction, i.e. $\vec{B}= B \hat{z}$, then the movement in $z$  will be constant, and we can ignore it. 

For the quantum case, the energy spectrum of the Landau problem is given by \cite{Landau}
\begin{equation}
E_n = \hbar \omega (n + \frac{1}{2}), \ \ \ n = 0, 1, 2,..., \ \ \omega=\frac{\vert e B \vert}{c},
\end{equation}
 these $E_n$ are  the well known Landau levels. However, for the Landau problem each level is infinitely degenerated. Therefore, we need an additional Hermitian operator, which commutes with $H$, to label the states.  If we choose the symmetric gauge, i.e. 

\begin{equation}
\vec{A}_s = \frac{B}{2}(-y,x),
\end{equation}
we can select the angular momentum in the $z$ direction, $L_z = x p_y - y p_x$, as the second operator.

In this gauge, the ground states are given by \cite{prlagen}
\begin{equation} \label{eq:edbasim}
\psi_{o,m}= \sqrt{\frac{1}{\pi m!} \left(\frac{e B}{2 \hbar c} \right)^{m+1}} (x + iy)^m e^{- \frac{e B }{4 \hbar c} \left( x^2 + y^2 \right)},
\end{equation}
where $m$ is a label for the angular momentum in the $z$ direction, such that
\begin{equation}
L_z \psi_{0,m} = \hbar m \psi_{0,m}.
\end{equation} 
In this case, we see that the wavefunction depends on the parameters space and the physical space $\vec{x}$. 



\section{The Quantum Metric Tensor of the Landau Problem} \label{sec:qmtoflanp}

The QMT, $G_{ij}$, is  useful to define a distance in the system's parameter space \cite{estreim}. If our quantum system depends on $n$ parameters, $\lambda_i$, the QMT is given by 
\begin{equation}\label{eq:qumetten}
G_{ij} = \Re e \left[ \left( \partial_i \psi  \right. \left. , \partial_j \psi \right)  \right] - \beta_i \beta_j, 
\end{equation}
where $\psi$ is the state of the system,  $\partial_i = \frac{\partial}{\partial \lambda_i}$ and
\begin{equation}
\beta_i =  - i \left( \psi, \partial_i \psi \right).
\end{equation}
 With this definition,  the corresponding distance will be \cite{estreim}
\begin{equation}
d l^2 = G_{ij} d\lambda_i d\lambda_j. 
\end{equation}

It is proved that the QMT is gauge invariant \cite{estreim}, nevertheless, this proof is not the most general. The demonstration   assumes some specific features of the phase difference caused by a gauge transformation. In order to show the gauge dependence of the QMT, we need to compute it in different gauges. The first calculations will be in the symmetric gauge.




\subsection{The Quantum Metric Tensor in the Symmetric Gauge } \label{sec:symgaug}
For the purpose of this paper,  it is sufficient to consider only the variation of  $B$,  therefore the parameter  space will be 1-dimensional,  with $\lambda_1 = B$, and setting $G_{BB}=G$ is appropriate. We will compute the QMT of the ground state with $m=0$, then, by using the state presented in (\ref{eq:edbasim}), the first term of the definition will be
\begin{equation} 
\Re e \left[ \left( \partial_B \psi, \partial_B \psi \right) \right]=\frac{1}{2 B},
\end{equation}
whereas
\begin{equation} \label{eq:betbpar}
\beta_B = 0, 
\end{equation}
therefore
\begin{equation} \label{eq:tenmequa}
G = \frac{1}{2 B}.
\end{equation}
\subsection{Comparison of  the Quantum Metric Tensor in different Gauges} \label{sec:section}
In order to prove the gauge dependence of the  QMT, we make the calculation in  different gauges.  It is known  \cite{sakurai} that when two gauges are related by 

\begin{equation} \label{eq:gautrpve}
\vec{A}_2 = \vec{A}_1 + \nabla \Lambda(\vec{\lambda}, \vec{x}),
\end{equation}
the corresponding wavefunctions obey 

\begin{equation} \label{eq:deffas}
\psi_2(\vec{\lambda}, \vec{x}) = \exp\left( i \frac{e }{\hbar c}  \Lambda \right) \psi_1(\vec{\lambda}, \vec{x}).
\end{equation}

According to  the theory \cite{estreim},  since the wavefunctions are related just by a change of phase, the QMTs should coincide. To explicitly show that this match does not always occur, we choose $\vec{A}_1$ as the symmetric gauge and
\begin{equation} \label{eq:delam}
\Lambda = g B xy.
\end{equation} 
This particular $\Lambda$ allows us to examine several gauges using $g$ as a parameter. In particular, when we set $g=\frac{1}{2}$, we obtain the Landau gauge $\vec{A}_L$ given by
\begin{equation}
\vec{A}_L= B \left( 0, x \right),
\end{equation} 
and with $g=0$ we recover the symmetric gauge.  In this case $\Lambda$ depends on the parameter $B$  and the physical space $(x,y)$.

Now, in equation  (\ref{eq:deffas}), we set $\psi_1$ as the ground state in the symmetric gauge, then the ground state with $m=0$ in the new gauge will be

\begin{equation} \label{eq:ebsicf}
\psi'_{0,0}(B,x,y)= \sqrt{\frac{e B}{2 \pi \hbar c}} \exp\left( - \frac{e B }{4 \hbar c} \left( x^2 + y^2 \right) \right) e^{ i \frac{ e g}{ \hbar c}  B x y}.
\end{equation}
From equation (\ref{eq:ebsicf}) and the definition of the QMT, we compute that
\begin{equation} \label{eq:tenmequa.1}
G'=\left( g^2 + \frac{1}{2} \right)\frac{1}{B}.
\end{equation}
The presence of $g$ in equation (\ref{eq:tenmequa.1}) clearly implies a gauge dependence. It is worth to notice, however, that the QMT diverges when $B \rightarrow 0$ for any  value of $g$.

\section{Real Gauge Invariant  Quantum Metric Tensor} \label{sec:reinqmt}

If  we perform a gauge transformation in the parameter space, given by
\begin{equation} \label{eq:gaugtransdgen}
\psi' = e^{i \alpha(\vec{\lambda}, \vec{x})} \psi,
\end{equation}
then  $\beta_i$  changes as
\begin{equation} \label{eq:trabetsgen}
\beta_i' = \beta_i + \left( \psi, (\partial_i\alpha) \psi \right).
\end{equation}
It has been assumed  that  the phase $\alpha$, as well as its derivatives, can be taken outside of the internal product. Therefore, we would be able to simplify equation (\ref{eq:trabetsgen}) to
\begin{equation} \label{eq:betsimpl}
\beta_i' = \beta_i + \partial_i \alpha.
\end{equation}
When the equation  (\ref{eq:betsimpl}) 
is valid, the tensor  (\ref{eq:qumetten}) is gauge invariant. This means that the QMT is gauge invariant when $\partial_i \alpha$ is independent of the measure of the internal product.

 However, some phases, and its derivatives, may depend on the physical space, $(x,y)$, or any other operators. See, for example, the phase in (\ref{eq:ebsicf}). In these cases equation (\ref{eq:trabetsgen}) cannot be simplified; thus, the tensor (\ref{eq:qumetten}) is no longer gauge invariant.  It is worth to notice that equation (\ref{eq:gautrpve}) and equation (\ref{eq:betsimpl}) seem to give the same transformation rule. However, in equation (\ref{eq:gautrpve}) the derivatives are computed respect to the coordinates, while in equation (\ref{eq:betsimpl}), one derives respect to the parameters $\lambda_i$.

Before constructing  the real gauge invariant QMT, we note that the equation (\ref{eq:qumetten}) can be written as
\begin{equation} \label{eq:qmtdcov}
G_{ij} = \Re e \left[ \left( \left( \partial_i - i \beta_i \right) \psi  \right. \left. , \left( \partial_j -i \beta_j \right) \psi \right)  \right], 
\end{equation}
or, in the coordinate representation
\begin{equation}
G_{ij} = \Re e \left[ \int d^3x \left( \partial_i + i \beta_i \right)\psi^* \left(\partial_j - i \beta_j \right)\psi \right],
\end{equation}
because $\beta_i$ is real.
Then, by looking the equation (\ref{eq:betsimpl}), we realize that $\beta_i$ transforms like a connection when $\alpha$ is independent of the internal product. This means that the QMT (\ref{eq:qmtdcov})  is  constructed with  covariant derivatives, using $\beta_i$ as the connection.  Nonetheless, in the general case $\beta_i$ transforms like (\ref{eq:trabetsgen}), and it cannot be used as the connection.

For constructing the gauge invariant QMT, we need a function $\Gamma_i$ that transforms like
\begin{equation} \label{eq:trnorgam}
\Gamma_i' = \Gamma_i +\partial_i \alpha, 
\end{equation}
when we perform a change of gauge given by (\ref{eq:gaugtransdgen}). 
With this new connection, the gauge invariant QMT will be
\begin{equation} \label{eq:tmqdercovgi}
G_{ij} = \Re e \left[ \left( \left( \partial_i - i \Gamma_i \right)\psi, \left( \partial_j - i \Gamma_j \right)\psi \right) \right], 
\end{equation}
or
\begin{equation} \label{eq:tmqenlarcnv}
G_{ij} = \Re e \left[ \int d^3x \left( \partial_i + i \Gamma_i \right)\psi^* \left(\partial_j - i \Gamma_j \right)\psi \right].
\end{equation}
In the equations (\ref{eq:tmqdercovgi}) and (\ref{eq:tmqenlarcnv}) we recognize de covariant derivative, $D_i$, given by
\begin{equation}
D_i = \partial_i - i \Gamma_i,
\end{equation}
which transforms like
\begin{equation} \label{eq:tradecov}
\left( D_i \psi \right)' = e^{i \alpha} D_i \psi,
\end{equation}
under a change of gauge.  Using the covariant derivative, the QMT takes the form
\begin{eqnarray} \label{eq:tmqcondcov}
G_{ij} = & \Re e \left( D_i \psi, D_j \psi \right) \nonumber \\  = & \Re e \left[ \int d^3x \left( D_i \psi \right)^* D_j \psi\right].
\end{eqnarray}
Equation (\ref{eq:tmqcondcov}) defines a gauge invariant QMT.  Since  equation (\ref{eq:tradecov}) is valid,  then (\ref{eq:tmqcondcov}) will always be gauge independent. However,  we need to find the correct connection that transforms like (\ref{eq:trnorgam}). 

The form of the new connection $\Gamma_i$ will depend on the specific problem to be analysed. $\Gamma_i$ must  reduce to $\beta_i$ in the case that  $ \partial_i \alpha$ can be taken outside of the internal product. Therefore, the new QMT must also reduce to (\ref{eq:qumetten}) when $\partial_i\alpha$ is independent of the measure of the internal product.  In the following section we present the  $\Gamma_B$ for the example studied in this paper.

\section{Gauge Invariant Quantum Metric Tensor for the Landau Problem} \label{sec:gaugeinvariant}

Continuing with the example presented in Section \ref{sec:section},  the new QMT is given by
\begin{equation} \label{eq:tmqpartlp}
G = \Re e \left( \left( \partial_B - i\Gamma_B \right) \psi , \left(\partial_B - i\Gamma_B \psi\right) \right).
\end{equation}
The fact that $\beta_B = 0$ in the usual case, suggests that we must set
\begin{equation}
\Gamma_B = 0,
\end{equation}
therefore, according to (\ref{eq:trnorgam}), and using that $\alpha= \frac{e g}{\hbar c} B xy$, we find 
\begin{equation}
\Gamma_B' = \dfrac{g e x y}{\hbar c},
\end{equation}
thus, under the transformation (\ref{eq:gaugtransdgen}), we get
\begin{equation} \label{eq:temqolpr}
G' = \Re e \left[ \int d^3x\left( \partial_B + i \dfrac{g e x y}{\hbar c} \right)(\psi')^*  \left(\partial_B - i \dfrac{g e x y}{\hbar c} \right) \psi' \right].
\end{equation}
Applying the equation (\ref{eq:temqolpr}) to the state given by (\ref{eq:ebsicf}), we obtain 
\begin{equation}
G'=\frac{1}{2 B},
\end{equation}
 for any gauge. That is, the QMT proposed in this paper is gauge independent.
 
\section{Conclusions} \label{sec:conclu}
We explicitly showed that the  QMT, introduced in \cite{estreim} depends on the gauge.  This dependence is directly related to the phase difference between the wavefunctions in different gauges: when the change of gauge introduces a  phase  whose derivatives $\partial_i \alpha$ can be taken outside of the internal product, the  QMT is invariant. However, when  general phases are considered, i. e. that depends on the internal product variables,  the QMT is gauge dependent. 

We also proposed a real gauge invariant  QMT by defining a new connection $\Gamma_i$ that transforms according to (\ref{eq:trnorgam}).  Despite the gauge independence, the connection $\Gamma_i$ was not explicitly given, and the form of $\Gamma_i$ will depend on the specific problem to be studied. In the example shown in this paper, i.e. the Landau Problem, we successfully proposed the correct $\Gamma_i$ for obtaining a gauge invariant QMT.

As it was pointed out  before, the QMT has several applications in physics. The fact that the QMT depends on the gauge can give rise to new studies in the topics that apply the QMT. For example, since the quantum fidelity is related  to the QMT, it is probable that the quantum fidelity also depends on the gauge.

Another important case is   the applicability of  the QMT for predicting quantum phase transitions. In the example studied in this paper, the gauge independent QMT, as well as the gauge dependent  QMT, diverges for the same value of the field $B$. This fact suggests that both QMTs are useful for predicting quantum phase transitions. However, the chosen gauge in this example is not the most general, and further studies  are necessary. 


\begin{acknowledgments}
This work was partially supported  by DGAPA-PAPIIT grant IN103716; CONACyT project 237503, and  scholarship 419420. 
We also wish to acknowledge   Unidad de Posgrado, UNAM for the support and the workshop "Academic Writing" during the preparation of this paper.
\end{acknowledgments}

\end{document}